\documentstyle[pra,aps,epsfig,preprint,array]{revtex}

\begin{document}


\title{Spin-Space Entanglement Transfer and Quantum Statistics}

\author{Y. Omar$^{1}$, N. Paunkovi\'{c}$^{1}$, S. Bose$^{1}$ and V.
Vedral$^{2}$}

\address{$^{1}$ Centre for Quantum Computation, Clarendon Laboratory,
University of Oxford, Oxford OX1 3PU, United Kingdom\\
$^{2}$ Optics Section, The Blackett Laboratory, Imperial College,
London SW7 2BZ, United Kingdom \\}

\date{4 March 2002}

\maketitle


\begin{abstract}

Both the topics of entanglement and particle statistics have
aroused enormous research interest since the advent of quantum
mechanics. Using two pairs of entangled particles we show that
indistinguishability enforces a transfer of entanglement from the
internal to the spatial degrees of freedom without any
interaction between these degrees of freedom. Moreover,
sub-ensembles selected by local measurements of the path will in
general have different amounts of entanglement in the internal
degrees of freedom depending on the statistics (either fermionic
or bosonic) of the particles
involved. \\

\end{abstract}



Since the advent of quantum mechanics, entanglement has been
identified as one of its most peculiar features
\cite{epr,scr,bell}. This "excess correlation" has recently become
an important resource in quantum information processing
\cite{qcom}. Entanglement is believed to be at the root of the
speed-up of quantum computers over their classical counterparts
\cite{shor}, and it also leads to an unconditionally secure
quantum cryptographic key exchange \cite{artur}.  Another
fundamental aspect of quantum physics, somewhat neglected in the
field of quantum information, is the distinction between two
different types of particles, fermions and bosons, manifested
through particle statistics (although see \cite{sougato} and for
fermions see \cite{lloyd,schliemann,antonio}). There are at first
sight two seemingly ``conflicting" views regarding the role of
indistinguishability and particle statistics in quantum
information processing. On the one hand, these two notions appear
to combine to offer ``natural" entanglement through forcing the
use of symmetrised and anti-symmetrised states (for bosons and
fermions respectively), and as we mentioned before, entanglement
is generally an advantage for quantum information processing
(although see \cite{herbut}). On the other hand,
indistinguishability prevents us from addressing the particles
separately which seems to be disadvantage in information
processing. In this article we analyze the role of
indistinguishability and particle statistics in a simple
information processing scenario.

Consider the following situation. Suppose that we have two pairs
of qubits (quantum two-level systems), each pair maximally
entangled in some internal degree of freedom. If the particles
carrying the qubits are of the same type -- say bosons -- but
distinguishable as a result of spatial separation, then we have
two units of entanglement (e-bits) in total. All of this
entanglement is in the internal degrees of freedom. If we now
consider bringing the particles close together and then
separating them again, without the internal degrees of freedom
ever interacting with the spatial ones, we should expect the whole
entanglement to remain in the internal degrees of freedom.
Surprisingly, as we demonstrate in this paper, a fraction of the
initial entanglement is transferred to the path degrees of
freedom of the particles. The fascinating implication is that the
transfer of entanglement is imposed by particle
indistinguishability and does not involve any controlled
operation between the internal and external degrees of freedom
(i.e. spin-path interaction), in contrast with the standard
entanglement swapping scheme \cite{entswap}. The prevalent
setting for local manipulations of entanglement in quantum
information processing either involves explicit interactions
between the internal degrees of freedom of two particles, or an
interaction of the internal degrees of freedom with some
apparatus. Here we introduce a completely different setting in
which particle paths are locally mixed without ANY interaction of
the internal degrees of freedom with anything else.

\begin{figure}[ht]
\begin{center}
\epsfig{file=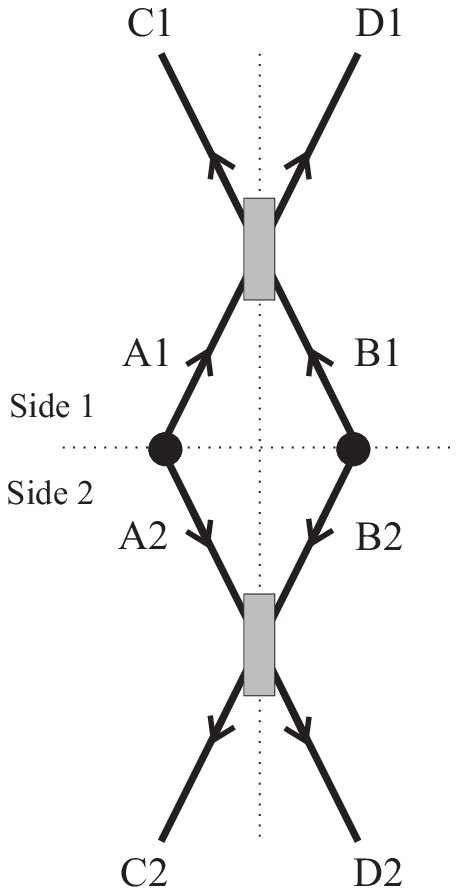, width=2in}
\end{center}
\caption{This figure presents our setup for spin-space
entanglement transfer. Each black circle represents a source of a
pair of particles maximally entangled in the internal degrees of
freedom (not explicitly shown in the figure). The rectangles
represent beam splitters.} \label{Fig. Setup}
\end{figure}

Now we turn to describing the exact details of our thought
experiment. Imagine the following setup, described in Fig.
\ref{Fig. Setup}. We have two pairs of identical particles, each
pair being maximally entangled in some internal degree of freedom,
e.g. the spin, or polarization. In our case, we consider systems
with spin one-half, or isomorphic to it. We assume that our setup
is symmetrical both horizontally and vertically, where the dotted
lines in Fig. \ref{Fig. Setup} show the axis of symmetry. We have
to ensure that particles arrive at the beam splitter at the same
time. The initial entanglement is between sides 1 and 2. In each
pair, the particles fly apart and meet a particle from the other
pair at a beam splitter. The paths on the left hand side are
labeled $A$ and $C$ respectively before and after the beam
splitter. Similarly, paths on the right hand side are labeled $B$
and $D$.

The output states of this setup represent particles in paths $C1$,
$D1$, $C2$ and $D2$ with a particular spin state (we note, for
instance, that we can have two particles in $C1$ and none in
$D1$). Now we show that, although the initial entanglement is only
in the internal degrees of freedom, in the final state some of the
entanglement has been transferred to the paths. We will refer to
this effect as the \textit{Spin-Space Entanglement Transfer} by
local actions only.

In order to calculate what happens in the above setup, we write
our initial state in the second quantization formalism:
\begin{equation} \label{Initial state}
\frac{1}{\sqrt{2}}(a_{A1\uparrow}^{\dag} a_{A2\downarrow}^{\dag}
\pm a_{A1\downarrow}^{\dag} a_{A2\uparrow}^{\dag})
\frac{1}{\sqrt{2}}(a_{B1\uparrow}^{\dag} a_{B2\downarrow}^{\dag}
\pm a_{B1\downarrow}^{\dag} a_{B2\uparrow}^{\dag})|0\rangle,
\end{equation}
where $|0\rangle$ is the vacuum state and, for instance,
$a_{A1\uparrow}^{\dag}$ is a creation operator describing a
particle in path A1 and with spin up. The positive and negative
signs in the above equation are necessary in order to take into
account all the possible initial states (the singlet and the
entangled triplet of spin). We restrict our attention to analyzing
one mode per particle only, but our results can be generalized to
any number of modes. Due to the symmetry of the problem we only
analyze two cases: when the two signs in equation (\ref{Initial
state}) are the same, the $(+,+)$ case, and when they are
different, the $(+,-)$ case. Note that initially there is no
left-right correlation between spin and space. This is because
there is no uncertainty in either the spin or space in the
initial state, so by measuring the spatial state one cannot gain
any information about the spin state (and vice-versa) in addition
to what we knew before the measurement.

\begin{figure}[ht]
\begin{center}
\epsfig{file=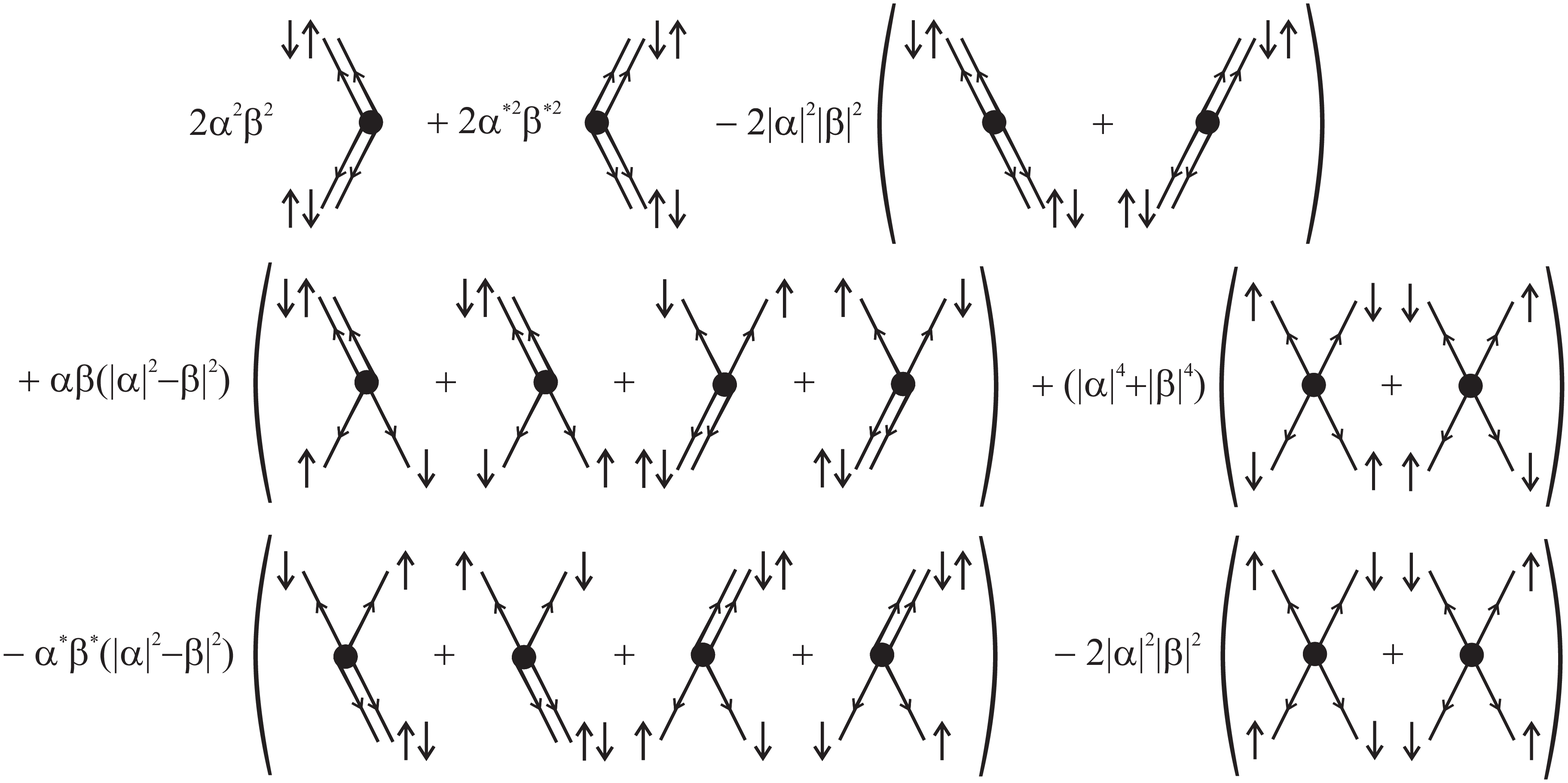, width=6in}
\end{center}
\caption{Spin $|S_z|=0$ component of the total output wave
function for the $(+,+)$ case, both for fermions and bosons.}
\label{Fig. CaseS0++}
\end{figure}

The operation of the beam splitter is described by any unitary
transformation in $U(2)$ \cite{loudon}. However, since the overall
phase factor has no relevance for entanglement, we can without any
loss of generality consider a transformation in $SU(2)$:
\begin{eqnarray}
U= \left[ \begin{array}{cc}
\alpha & \beta \\
-\beta^{*} & \alpha^{*} \\
\end{array} \right],
\end{eqnarray}
where $|\alpha|^{2}+|\beta|^{2}=1$. Since we consider entanglement
only between sides $1$ and $2$, the beam splitters in fact perform
local unitary operations. Hence they cannot change the total
entanglement present initially. Also, they only affect the spatial
degrees of freedom and are not intrinsically dependent on spin (or
polarization). Therefore they are incapable of swapping
entanglement from spin (polarization) to space by performing a
controlled not operation in the usual fashion \cite{entswap}.
Although the transformation law will be the same for fermions and
bosons, they obey different statistics which is why there will be
an observable difference in their behaviour in our experiment. For
fermions we have the following anti-commutation relation:
\begin{equation}
[a_{i}^{\dag},a_{j}^{\dag}]_{+}=0,
\end{equation}
while for bosons we have the commutation relation:
\begin{equation}
[a_{i}^{\dag},a_{j}^{\dag}]_{-}=0,
\end{equation}
where $i$ and $j$ are sets of labels. Figs. \ref{Fig. CaseS0++} to
\ref{Fig. CaseS1+B} present diagramatically the output states for
both fermions and bosons. For instance, the first diagram in Fig.
\ref{Fig. CaseS1+F} represents the following term:
\begin{equation}
( |\alpha|^{2} + |\beta|^{2})^{2} \, (a_{C1\uparrow}^{\dag}
a_{D1\uparrow}^{\dag}a_{C2\downarrow}^{\dag}
a_{D2\downarrow}^{\dag})|0\rangle.
\end{equation}

Note that for each output pair, i.e. both on sides $1$ and $2$,
the total spin (or polarization) $S$ can take the values $0$ or
$1$. If we consider, without any loss of generality, that the spin
is aligned with the $z$ axis, then $|S_z|$ -- the absolute value
of the projection of \textbf{S} along \textit{z} -- can also only
take the values $0$ or $1$. We can then divide the total output
wave function into these two components, where the spins of the
particles in each output pair are respectively anti-aligned or
aligned along $z$.

\textbf{$|S_z|=0$ component:} there is no difference between
fermions and bosons (bearing in mind that the corresponding
operators obey different commutation relations). However, there
is a difference between the $(+,+)$ case, where we have all
possible output terms (see Fig. \ref{Fig. CaseS0++}), and the
$(+,-)$ case, where some terms never appear (see Fig. \ref{Fig.
CaseS0+-}).

\textbf{$|S_z|=1$ component:} there is a difference between the
output states for fermions (see Fig. \ref{Fig. CaseS1+F}) and
bosons (see Fig. \ref{Fig. CaseS1+B}). For both types of
particles, the $(+,-)$ case will only introduce a phase
difference in some terms.

As a consequence of applying only local unitary operations, the
total output wave function should have also two e-bits of
entanglement. For clarity, let us consider for the rest of the
paper the particular case of $50/50$ beam splitters
($\alpha=1/\sqrt{2}, \beta=-i/\sqrt{2}$). To illustrate the
spin-space entanglement transfer effect, we look at the $(+,+)$
case for fermions (Figs. \ref{Fig. CaseS0++} and \ref{Fig.
CaseS1+F}). Here, it is clear that the $|S_z|=1$ terms give one
e-bit of entanglement, solely in the internal degrees of freedom,
as the path states are identical. The $|S_z|=0$ case gives the
other e-bit of entanglement, but this time involving both the
internal and external degrees of freedom. \textit{Thus we have
spin-space entanglement transfer, without any controlled
operation between spin and space.}

We now show how we can extract space-only entanglement from the
total wave function by doing particular measurements on the
internal degrees of freedom without revealing any knowledge about
the external ones (this is perfectly allowed by quantum mechanics
and can be accomplished by passing the particles on each side
through a cavity which extends over both the left and the right
paths). For example, we can measure the total spin \textbf{S} on
both sides (1 and 2) along the $x$ axis and then select the
$S_x=0$ results. For fermions, the entire wave function is then
projected onto:
\begin{equation}
\frac{1}{\sqrt{2}} \left[ \frac{1}{\sqrt{2}} \left( |L\rangle_{1}
+ |R\rangle_{1} \right) \frac{1}{\sqrt{2}} \left( |L\rangle_{2} +
|R\rangle_{2} \right) -
\frac{1}{\sqrt{2}}|A\rangle_{1}\frac{1}{\sqrt{2}}|A\rangle_{2}
\right],
\end{equation}
where $|L\rangle_{1,2}$ means left bunching of the particles,
respectively for sides $1$ and $2$, $|R\rangle_{1,2}$ right
bunching and $|A\rangle_{1,2}$ represents anti-bunching
(unormalized state). The bosonic counterpart of the above state
is:
\begin{equation}
\frac{1}{\sqrt{2}} \left[ \frac{1}{\sqrt{2}} \left( |L\rangle_{1}
+ |R\rangle_{1} \right) \frac{1}{\sqrt{2}} \left( |L\rangle_{2} +
|R\rangle_{2} \right) +
\frac{1}{\sqrt{2}}|A\rangle_{1}\frac{1}{\sqrt{2}}|A\rangle_{2}
\right].
\end{equation}
Both these states have 1 e-bit of entanglement in space and the
same outcome probability of $1/2$. Note that since these two
states are orthogonal they can be perfectly discriminated,
offering an operational way of distinguishing fermions and bosons.

If on the other hand we measure the spatial components of the
total wave function, we will find different amounts of
entanglement in the internal degrees of freedom of fermions and
bosons. For instance, if we select the anti-bunching results, we
will obtain the following state for fermions:
\begin{eqnarray}
\frac{1}{\sqrt{3}} \left[ \frac{1}{\sqrt{2}}
(a_{C1\uparrow}^{\dag}a_{D1\downarrow}^{\dag} +
a_{C1\downarrow}^{\dag} a_{D1\uparrow}^{\dag}) \frac{1}{\sqrt{2}}
(a_{C2\uparrow}^{\dag}a_{D2\downarrow}^{\dag} +
a_{C2\downarrow}^{\dag} a_{D2\uparrow}^{\dag}) \right. \nonumber \\
\left. - (a_{C1\uparrow}^{\dag}
a_{D1\uparrow}^{\dag}a_{C2\downarrow}^{\dag}
a_{D2\downarrow}^{\dag}) - (a_{C1\downarrow}^{\dag}
a_{D1\downarrow}^{\dag}a_{C2\uparrow}^{\dag}
a_{D2\uparrow}^{\dag}) \right] |0\rangle,
\end{eqnarray}
with an outcome probability of $2/3$ and $\log_{2}3$ units of
entanglement, whereas for bosons we will get:
\begin{equation}
\frac{1}{\sqrt{2}} (a_{C1\uparrow}^{\dag}a_{D1\downarrow}^{\dag} -
a_{C1\downarrow}^{\dag} a_{D1\uparrow}^{\dag}) \frac{1}{\sqrt{2}}
(a_{C2\uparrow}^{\dag}a_{D2\downarrow}^{\dag} -
a_{C2\downarrow}^{\dag} a_{D2\uparrow}^{\dag}) |0\rangle,
\end{equation}
with probability $1/3$ and $0$ units of entanglement.

If we select the bunching results, for fermions we will obtain
the state:
\begin{equation}
\frac{1}{\sqrt{2}} (a_{C1\uparrow}^{\dag}a_{C1\downarrow}^{\dag} +
a_{D1\uparrow}^{\dag} a_{D1\downarrow}^{\dag}) \frac{1}{\sqrt{2}}
(a_{C2\uparrow}^{\dag}a_{C2\downarrow}^{\dag} +
a_{D2\uparrow}^{\dag} a_{D2\downarrow}^{\dag}) |0\rangle,
\end{equation}
with an outcome probability of $1/3$ and $0$ units of
entanglement, and for bosons we will get:
\begin{eqnarray}
\frac{1}{\sqrt{3}} \left[ \frac{1}{\sqrt{2}}
(a_{C1\uparrow}^{\dag}a_{C1\downarrow}^{\dag} +
a_{D1\uparrow}^{\dag} a_{D1\downarrow}^{\dag}) \frac{1}{\sqrt{2}}
(a_{C2\uparrow}^{\dag}a_{C2\downarrow}^{\dag} +
a_{D2\uparrow}^{\dag} a_{D2\downarrow}^{\dag}) \right. \nonumber
\\ + \frac{1}{2} (a_{C1\uparrow}^{\dag}a_{C1\uparrow}^{\dag} +
a_{D1\uparrow}^{\dag} a_{D1\uparrow}^{\dag})
(a_{C2\downarrow}^{\dag}a_{C2\downarrow}^{\dag} +
a_{D2\downarrow}^{\dag} a_{D2\downarrow}^{\dag}) \nonumber
\\ \left. + \frac{1}{2} (a_{C1\downarrow}^{\dag}a_{C1\downarrow}^{\dag} +
a_{D1\downarrow}^{\dag} a_{D1\downarrow}^{\dag})
(a_{C2\uparrow}^{\dag}a_{C2\uparrow}^{\dag} +
a_{D2\uparrow}^{\dag} a_{D2\uparrow}^{\dag}) \right] |0\rangle,
\end{eqnarray}
with probability $2/3$ and $\log_{2} 3$ units of entanglement. We
observe that for a given path selection one type of particles
exhibits some entanglement in the internal degrees of freedom,
whereas the other exhibits none. In other words, under the
\textit{same} situation, fermions and bosons show a difference in
their information processing behaviour. Moreover, measuring this
degree of entanglement in the internal degrees of freedom could
thus also be an operational way of distinguishing between
fermions and bosons.


In this article we have shown that it is possible to transfer
entanglement from the internal to the spatial degrees of freedom
through local actions using only the effects of particle
indistinguishability and quantum statistics, without any
interaction between the spin and the path. Moreover,
sub-ensembles selected by local measurements of the path will in
general have different amounts of entanglement in the internal
degrees of freedom depending on the statistics (either fermionic
or bosonic) of the particles involved. This establishes a
connection between two fundamental notions of quantum physics:
entanglement and particle statistics. We intend to present a more
detailed and systematic analysis of this setup in a subsequent
longer work.

Our analysis suggests further investigation of the consequences
and applications of particle statistics in quantum information
processing. For example, in some protocols using spin-space
entanglement the statistical effects make it unnecessary to have
controlled operations, such as using polarization-dependent beam
splitters \cite{us}. Other types of statistics (e.g. anyons) can
similarly be addressed within our framework. Recent experiments
such as \cite{kwiat,yamamoto} suggest that it would be possible
to test our results in the near future.

Y.O. acknowledges support from Funda\c{c}\~{a}o para a Ci\^{e}ncia
e a Tecnologia from Portugal. N.P. thanks Elsag S.p.A. for
finantial support. V.V. acknowledges support from Hewlett-Packard
company, EPSRC and the European Union project EQUIP.


\begin{figure}[ht]
\begin{center}
\epsfig{file=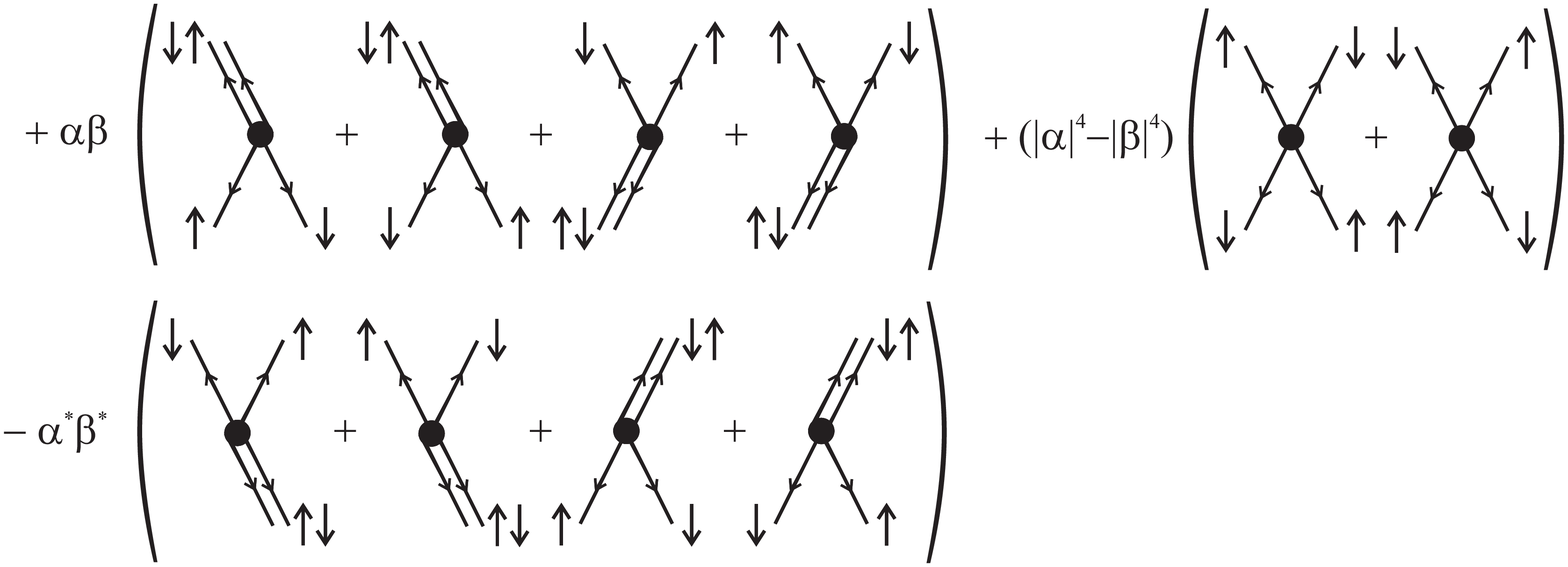, width=6in}
\end{center}
\caption{Spin $|S_z|=0$ component of the total output wave
function for the $(+,-)$ case, both for fermions and bosons.}
\label{Fig. CaseS0+-}
\end{figure}

\begin{figure}[ht]
\begin{center}
\epsfig{file=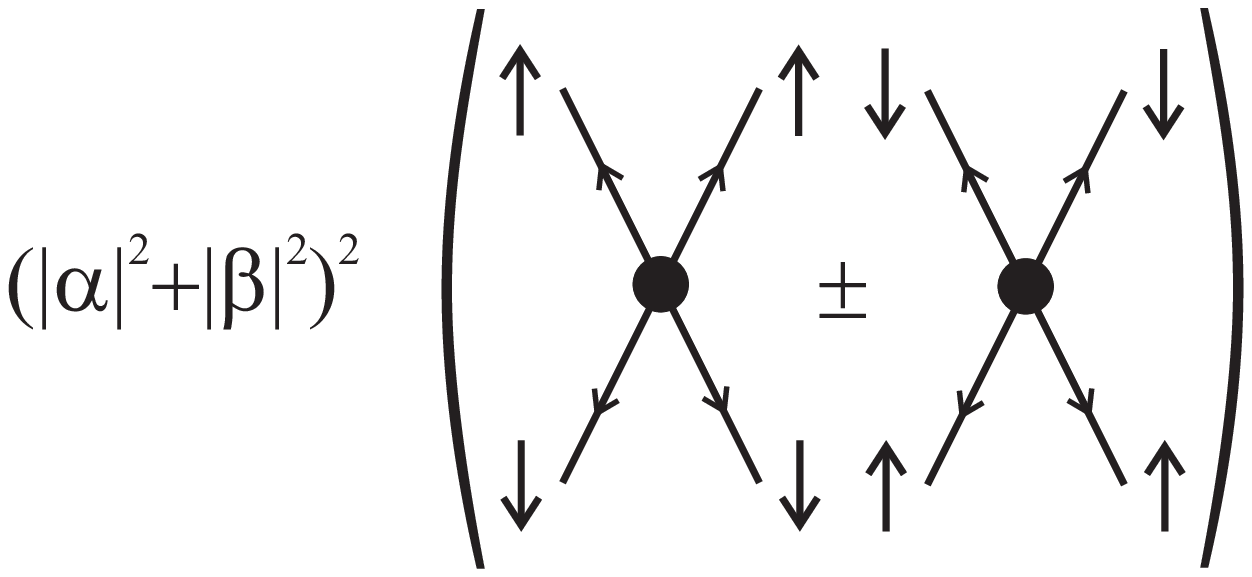, width=2.5in}
\end{center}
\caption{Spin $|S_z|=1$ component of the total output wave
function for the $(+,\pm)$ cases, for fermions.} \label{Fig.
CaseS1+F}
\end{figure}

\begin{figure}[ht]
\begin{center}
\epsfig{file=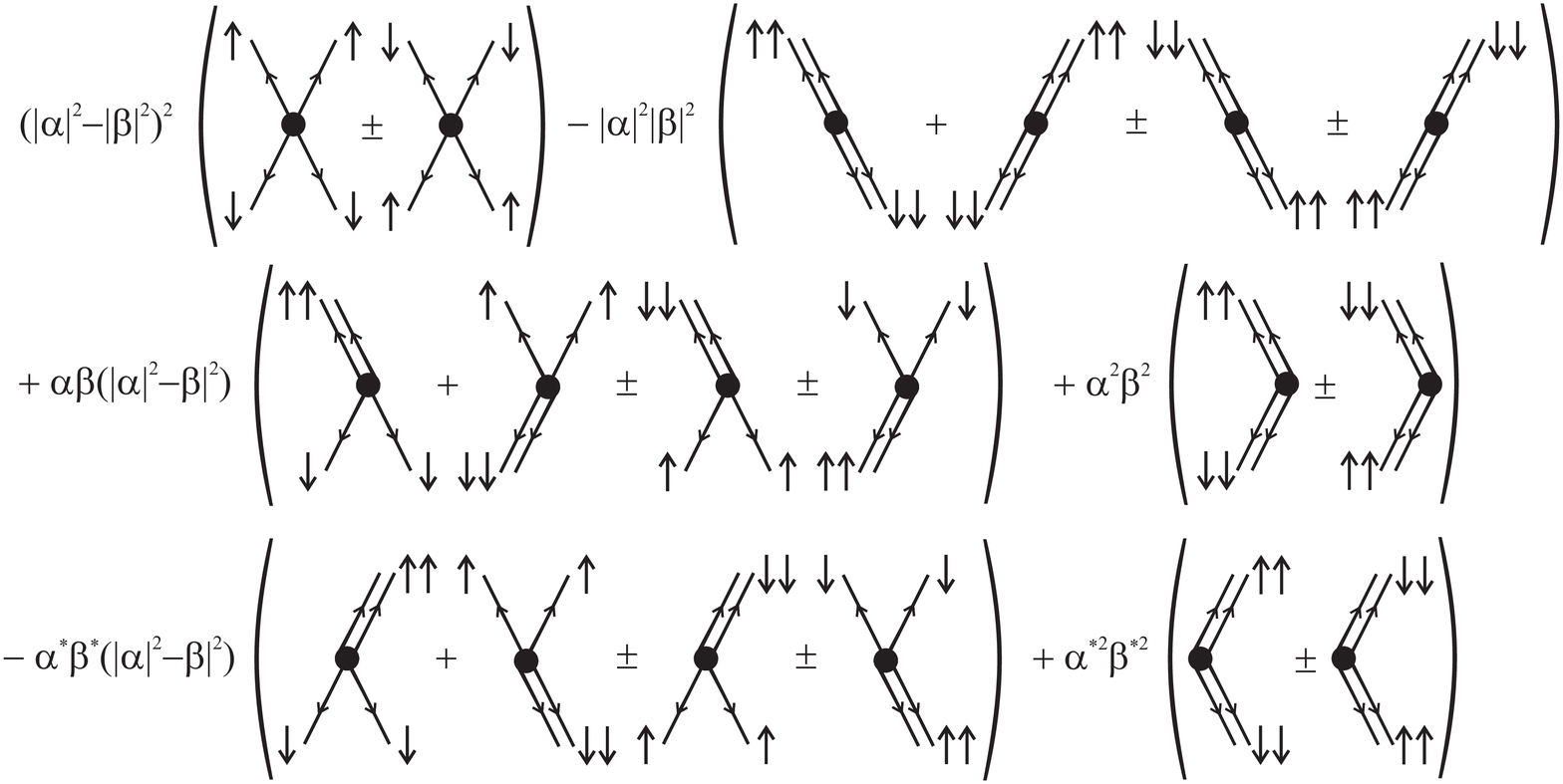, width=6in}
\end{center}
\caption{Spin $|S_z|=1$ component of the total output wave
function for the $(+,\pm)$ cases, for bosons.} \label{Fig.
CaseS1+B}
\end{figure}


\end{document}